\documentclass[aps,pra,twocolumn,superscriptaddress,floatfix]{revtex4}

\usepackage{bm}
\usepackage[colorlinks=true, citecolor=blue, urlcolor=blue]{hyperref}\usepackage{epsf,graphics,graphicx}
\usepackage{amsmath}
\usepackage{latexsym}
\usepackage{amssymb}
\usepackage{graphicx}
\usepackage{color}
\usepackage{soul,xcolor}
\setstcolor{red}
\usepackage{ulem}
\usepackage{cancel}
\usepackage{tikz}
\usepackage[export]{adjustbox}
\usepackage{subcaption}
\usepackage{braket}

\newcommand{\beq}{\begin{equation}}
\newcommand{\eeq}{\end{equation}}

\def\bea{\begin{eqnarray}}
\def\eea{\end{eqnarray}}
\def\ba{\begin{array}}
\def\ea{\end{array}}

\usepackage{epsfig}

\input{epsf}

\begin{document}
\title{Anomalous Hall transport in tilted multi-Weyl semimetals}
\author{Anirudha Menon}
	\affiliation{Department of Physics, University of California, Davis, California 95616, USA} \email{amenon@ucdavis.edu}
	\author{Banasri Basu}
	\affiliation{Physics and Applied Mathematics Unit, Indian Statistical Institute, Kolkata 700108, India}
	\email{sribbasu@gmail.com}

\date{\today}

\begin{abstract}
We study the effect of a perpendicular magnetic field $\mathbf{B}$ on a multinode Weyl semimetal (mWSM) of arbitrary integer monopole charge $n$, with the two Weyl multinodes separated in $\mathbf{k}$-space. Besides type-I mWSMs, there exist type-II mWSMs which are characterized by the tilted minimal dispersion for low-energy excitations; the Weyl points in type-II mWSMs are still protected crossings but appear at the contact of the electron and hole pockets, after the Lifshitz transition. We find that the presence of a perpendicular magnetic field quantizes the occupation pockets due to the presence of Fermi tubes. In this theory, the Hilbert space is spanned by a set of $n$ chiral degenerate ground states, and a countably infinite number of particle-hole symmetric Landau levels.  We calculate the Hall conductivity for the tilt-symmetric case of type-I mWSM using the Kubo formula, in the zero-frequency (DC) limit, and recover the well-known vacuum contribution. We compute the Fermi surface corrections and show that the expression generalizes from the formula for elementary ($n=1$) type-I WSMs. We derive an expression for the type-II mWSM Hall conductivity, which is bounded by a Landau level cutoff introduced on physical grounds. Interestingly, we find that the anomalous vacuum Hall conductivity is vanishing in the type-II phase at all temperatures. The corresponding thermal Hall and Nernst conductivities are evaluated and characterized for both phases. The qualitative and quantitative observations presented here may serve in the characterization of generic mWSMs of both types.
\end{abstract}

\maketitle

\section{Introduction}
Weyl fermion like quasi-particles have been proposed to exist in condensed matter systems in recent years \cite{new3,new5,new6,new7,arp1,arp2}, and have been realized through tabletop experiments \cite{new4,5,6,7}. They find theoretical generalizations to a class of quasi-particle excitations characterized by topological invariants, manifested in materials called Weyl semimetals \cite{11,12,8,9,10}. Topological materials can form exceptions \cite{new11,new12} to the famous Neumann-Wigner no-crossing theorem \cite{new10} of Bloch bands, which states that there can be no level-crossing if two bands share the same symmetry.\\

A generic Weyl semimetal (WSM) is considered to be a topological quantum system, with two {\bf k}-space monopoles having opposite topological charge, and a gapless spectrum \cite{new3,new4,new5,new6,new7,11,12}. The integer charged monopoles occur in pairs and act as a source or sink for the Berry flux, i.e., the surface integral of the ${\bf U(1)}$ Berry 2-form \cite{new21,KTA, Tong}. These monopoles constitute the Weyl nodes, which are the points at which the valence and conduction bands touch. The key to Weyl physics lies in the fact that the Fermi surface should be sufficiently close to the Weyl points. \\

In principle, Weyl nodes of opposite chirality merge and annihilate in pairs, while those of the same chirality can merge to form nodes of larger topological charge, which are stable provided that there is a point group symmetry protecting the merger \cite{12}. The k-space merger of Weyl points with the same chirality produces a new type of WSM referred to as multi-Weyl semimetal. One can then understand the multi-Weyl semimetal (mWSM) as a robust state of mergers of $m$ like-chirality unit charge monopoles or elementary mWSMs and it's worth noting that the only allowed values of monopole charge leading to point group symmetry protected mWSMs are $n=1,2,3$ \cite{12}. The low energy description involves minimal models for a mWSM, which require either inversion or time-reversal symmetry to be broken \cite{14,15,16}, and give rise to Dirac-like dispersions \cite{new14,new15} at the Weyl nodes along a symmetry direction. The remaining directions contribute to the energy non-linearly, dictated by the monopole charge, leading to anisotropy \cite{8,10,KTA}. In the low energy description, a Lorentz symmetry violating tilt ($C$) term can be used to induce a new phase of elementary WSMs known as type-II WSMs \cite{WSMII,FP1,FP2}, with $C \gg v$, $v$ being the Fermi velocity. Materials with this band structure have quasi-particle pocket Fermi surfaces at charge neutrality, compared to the point like Fermi surfaces for type-I WSMs ($C \ll v$). This construction can be extended \cite{mWSMII} to WSMs of arbitrary winding number and can be used to probe the properties of type-II mWSMs sufficiently far away from the topological Lifshitz transition separating the two phases. Both phases of elementary Weyl semimetals host novel surface states called Fermi Arcs \cite{new19}, a concept which extends smoothly to mWSMs. \\

It had been theorized that both phases of Weyl semimetals host \cite{new55,new57,new58,new59,new62} the condensed matter equivalent of the Adler-Bell-Jackiw anomaly (also known as chiral anomaly) \cite{new54,new56} which is the violation of independent current conservations for left and right-handed Weyl fermions. Elementary WSMs have been examined both theoretically and experimentally in the contexts of magneto-optical transport \cite{US}, negative magneto-resistance \cite{NMR}, and a plethora of related effects \cite{Tchoumakov16,yu16,udagawa16,bitan16,trescher18,23,new53,new60,new61,new63,new64,new65,
new66,new67,new69}. Recent theoretical reports claim $\textrm{SiSr}_2$ \cite{9} and $\textrm{HgCr}_2\textrm{Se}_4$ \cite{11,12} as possible candidates for mWSMs with monopole charge $n=2$. The dispersion anisotropy in mWSMs coupled with spin-momentum locking \cite{8} has the potential to give rise to unique quantum effects and transport signatures \cite{huang17,new82,new83,new84,new86}.\\

The classical Hall effect \cite{Hall1879} and its anomalous variant \cite{new1}, discovered by E. Hall in the nineteenth century, have since found fascinating quantum versions \cite{new2,Tong} of significant interest to the condensed matter community. The quantum anomalous Hall effect (QAHE or AHE) refers to the contribution to Hall conductivity from spin-orbit coupling in ferromagnetic material \cite{new2, zyuzin16}. This effect is usually dependent on the magnetization of the material and comes in two flavors: {\it intrinsic} or scattering between bands, and {\it extrinsic} or impurity scattering. Hall physics and its connection to the geometric phase has been studied extensively for elementary WSMs an we discuss some key discoveries in what follows. The {\it zero-mode} or {\it vacuum contribution} to the AHE, which is ground state scattering contribution, turns out to be universal \cite{gorbar14} in the type-I phase of WSMs (i.e., independent of temperature, chemical potential, and tilt) and is known in the literature as the ``topological" term. In the type-II phase it has been shown to be non-universal in the minimal model \cite{zyuzin16}. AHE and the thermal Hall and Nernst coefficients have been calculated in the minimal model of type-I and type-II WSMs \cite{new41,zyuzin16,new44,mukher18}, in a perpendicular magnetic field \cite{21,gorbar14} (no tilt), and in lattice models \cite{new43,gorbar17}, with subsequent experimental validation \cite{new46,new47,new48}. The thermo-electric effect has been studied semi-classically in WSMs \cite{chen16,park17,new42} and additionally including of Berry curvature effects \cite{new52} using the Boltzmann formalism.\\

Motivated by the prospect of rich and novel physics, in this paper we study the intrinsic AHE of both types of mWSMs in a perpendicular magnetic field. We find that under the influence of a perpendicular magnetic field, the continuum band structure of the minimal model mWSM splits into discrete copies of one dimensional dispersions in a plane perpendicular to the direction of quantizing field. The degenerate ground states of this system are chiral, as a recent experimental study \cite{24} indicates, leading to interesting consequences on Hall transport. In particular, we pay close attention to the anomalous vacuum contribution, known to be universal in the absence of a magnetic field in the type-I $n=1$ mWSM case, and non-universal in the type-II phase \cite{zyuzin16}. Additionally, Luttinger's phenomenological theory of transport \cite{Lutt}, the Wiedemann-Franz law, and Mott's rule \cite{14,US} permit the study of the thermal Hall and Nernst conductivities. \\  

Recently the effects of a perpendicular magnetic field on the anomalous Hall conductivity has been examined in $n=1$ type-I mWSMs in the minimal model \cite{gorbar14,21}, but no such attempts have been made for the type-II phase or indeed general mWSMs (besides the topological vacuum contribution in the type-I mWSM phase \cite{huang17}). In this manuscript, we pursue these results in the minimal models of both phases of mWSMs. The rest of the paper is organized as  follows: In Section II we discuss time-reversal symmetry breaking model Hamiltonian for the mWSM in a perpendicular magnetic field and the corresponding dispersion relations. Section III narrates the underlying physics of the band structure due to Landau levels for the type-I and type-II mWSMs where the limitations of the minimal model are explained. Section IV deals with computation of the AH conductivity for both types of mWSMs, including the respective Fermi surface corrections to the vacuum conductivity. We explore the implication of the chiral zero modes induced by the transition of the mWSM electronic structure by the quantizing field on the off-diagonal transport properties of the system. Additionally, results for type-I and type-II anomalous thermal Hall and Nernst conductivities are explained. In Section V we discuss the physics leading to the annihilation of the vacuum contribution for the type-II mWSM phase in the $T \rightarrow 0$ limit, and then we demonstrate that it is vanishing at all temperatures in this phase. Section VI sums up our findings. \\

\section{The model}

 Our discussion begins with the minimal Hamiltonian for a pair of multi-Weyl nodes given by \cite{KTA,TNAMBB} 

\begin{align}
H_{n}^{s}=\hbar C_{s}(k_z-sQ)+s\hbar \alpha_n \bm\sigma\cdot{\bm n}_p,
\label{1}
\end{align}

where $s=\pm$ characterizes the Weyl point (WP), $C_s$ is the tilt parameter, which can be different for each node, in principle. Here, ${\bm n}_p=\frac{1}{\hbar} \left[p_{\perp}^n\cos(n\phi_p), p_{\perp}^n\sin(n\phi_p), \frac{v(p_z- s\hbar Q)}{\alpha_n}   \right]$, $p_{\perp} =\sqrt{p_x^2+p_y^2}$, $\bm \sigma$ is the vectorized Pauli matrix, $v$ denotes the Fermi velocity in the absence of tilt, and $n$ is the monopole charge. This Hamiltonian has mW nodes separated by $2Q$ along $\bm e_z$, which is the unit vector along the z-direction in momentum space, and $\alpha_n$ constitutes the dimensionally consistent generalization of the Fermi velocity in the $k_x-k_y$ plane. For a derivation of this model for $n=1,2,3$ form the corresponding lattice model, see \cite{App} section A.\\

\begin{widetext}
	
	\begin{figure}[h]
		\centering
		\fbox{\includegraphics[scale = .8]{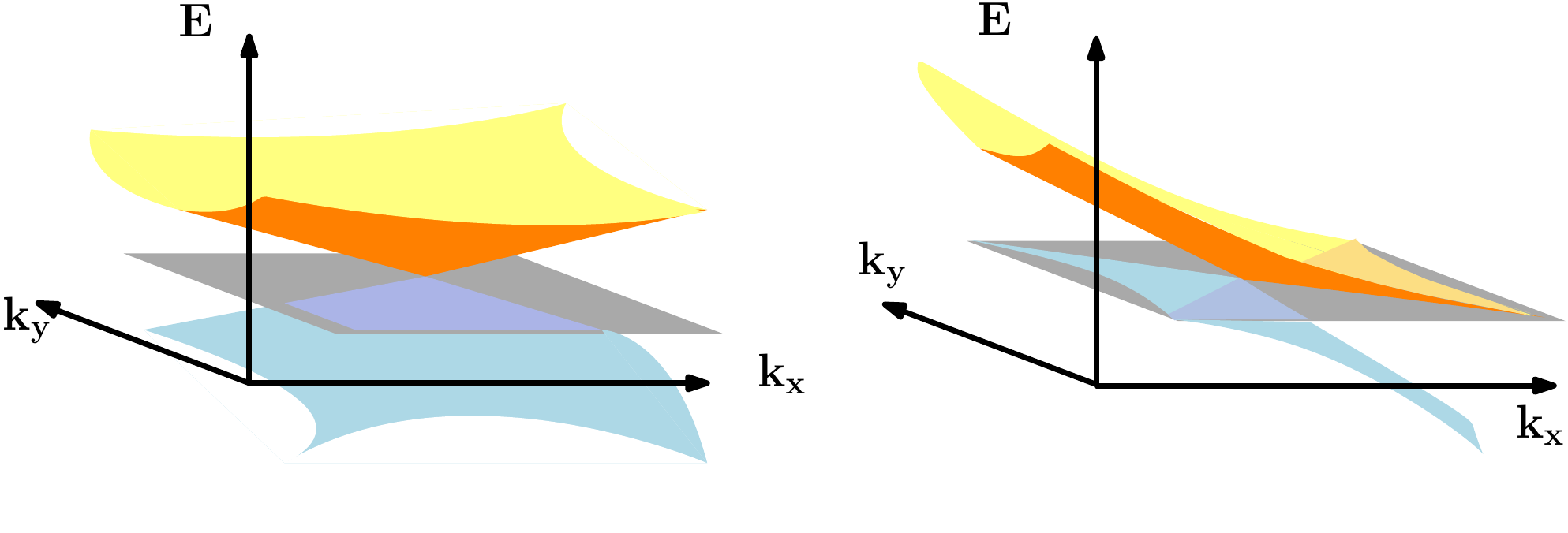}}
		\caption{Left panel: Spectrum of a type-I mWSM in the absence of a magnetic field. Right panel: Tilted type-II mWSM spectrum with $B = 0$, showing the formation of electron and hole pockets characteristic of this phase. } 
	\end{figure}

	\begin{figure}[h]
		\centering
		\fbox{\includegraphics[scale = .8]{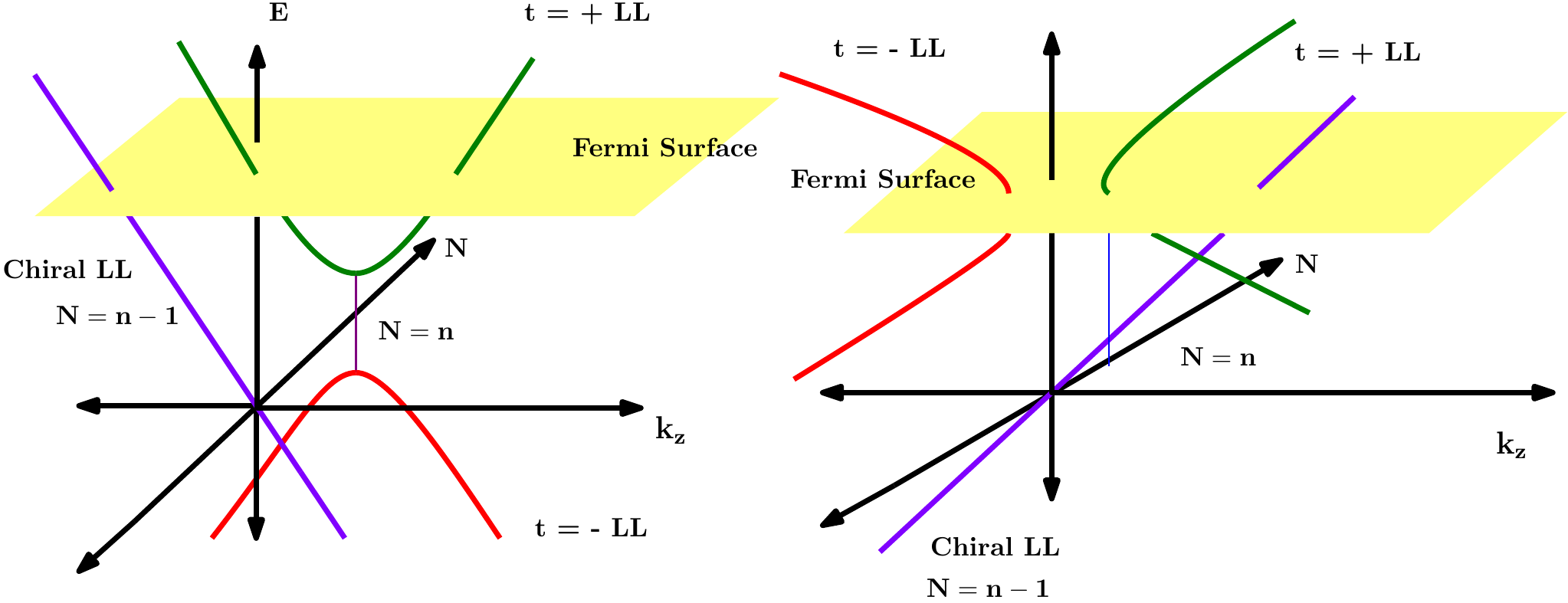}}
		\caption{ Left panel: Type-I mWSM spectrum in the presence of a perpendicular magnetic field. The achiral valence and conduction bands are in green and red respectively, with the Fermi surface shown in yellow, while the chiral bands are depicted in violet. Right panel: Type-II mWSM with tilted bands showing that every LL $N$ is partially occupied.}
\end{figure}	

\end{widetext}

The dispersion of the $s = +$ ($Q = 0$, $C_+ = C > 0$) node of a mWSM is given by \cite{10, KTA}

\begin{eqnarray} \label{CB1}
E_t({\mathbf k}) = Ck_z + tv\sqrt{k_z^2+ \gamma^2(k_x^2 +k_y^2)^n} , 
\end{eqnarray}

The quantum number $t = +(-)$ denotes the conduction (valence) bands and  
 $\gamma$ is a dimensionful constant (given by $\gamma  = \alpha_n /v$), where  $v$ is the Fermi velocity. We proceed by introducing a magnetic field perpendicular to the $x-y$ plane in the Landau gauge: $\bm A=x B {\hat{\bm y}}$ such that $\bm B=\bm {\nabla} \times \bm A=-B {\hat{\bm z}}.$ The Hamiltonian is presented in the compact matrix form:

\begin{equation}\label{ham6} 
H^n_{s,B} = \left[ \begin{array}{cc}  (C_s + s v) z &  i^n s \frac{\alpha_n}{\ell_B^n}(\sqrt{2} a^\dagger)^n \\ \\
  -  i^n s \frac{\alpha_n}{\ell_B^n}( \sqrt{2} a)^n    & (C_s - s v) z  
  \end{array} \right] ,
\end{equation}

with the introduction of the ladder operators $a(a^{\dagger})$, following the Pierel's substitution $p_i \rightarrow p_i - eA_i$, and setting $c = 1$, $\hbar =1$, $z = k_z -sQ$. The spectrum $E_N^{t,s}$ and eigenstates of the mW Hamiltonian for the Landau levels $N \geq n$ are given by \cite{App} (section B)

\begin{align}\label{en2}
E_N^{t,s} &=& C_s z+t v \sqrt{z^2+ {^NP_n} \Omega^2} \nonumber \\
&=& \frac{C_s}{v} F_0 + tF_N,
\end{align}

\begin{align}\label{vector}
\mid N,s,t,k_z\rangle  &=  \left[  \begin{array}{cc} C_{\uparrow,s,t,k_z,N} |N\rangle \\ \\ 
  C_{\downarrow,s,t,k_z,N} \mid N-n\rangle \end{array} \right] ,
\end{align}

where, $\Omega$ represents the LL spacing \cite{App} (section B), $F_0 \equiv v(k_z-sQ)$, and $F_N \equiv v\sqrt{(k_z-sQ)^2 + \Omega^2 {}^NP_n}$. The eigenstates have been unit-normalized, and the quantum number $t = +(-)$ denotes the conduction (valence) bands, equivalently called the valley degrees of freedom. Here, the dependence of the wavefunction on the plane waves in the $y$ and $z$ directions have been suppressed for brevity, and since the dispersion has no dependence on $k_y$, it is macroscopically degenerate. The explicit expressions for the coefficients in eqn. (\ref{vector}) are given by, 

\begin{eqnarray}\label{a11-12}
 	C_{\uparrow,s,t,k_z,N} &=& \frac{1}{\sqrt{2}}\left[ 1+\frac{E_z}{E_{N}^t-E_c}\right]^{1/2} \delta_{s,+} \nonumber \\
 	&+&  
 	\frac{1}{\sqrt{2}}\left[ 1 -\frac{E_z}{E_{N}^t-E_c}\right]^{1/2} \delta_{s,-} \nonumber
 	\end{eqnarray}
 	\begin{eqnarray}
 	C_{\downarrow,s,t,k_z,N} &= & \frac{-i^n t}{\sqrt{2}}\left[ 1 -\frac{E_z}{E_{N}^t-E_c}\right]^{1/2} \delta_{s,+} \nonumber \\
 	&+&  
 	\frac{i^n t}{\sqrt{2}}\left[ 1 +\frac{E_z}{E_{N}^t-E_c}\right]^{1/2} \delta_{s,-}.
 	\end{eqnarray}

 The Landau levels, characterized by $N$ in the elementary WSM case, now generalize to $^NP_n$, where $P$ is the permutation operator defined as $^NP_n = \frac{N!}{(N-n)!}$. The modes with $0 \leq N \leq n-1$ have the form

\bea
|N=n-q,s,k_z, s\cdot t =+ \rangle &= \begin{bmatrix}
                       |n-q\rangle \\ 0
                      \end{bmatrix},  \nonumber \\
1 \leq q \leq n , & q \; \epsilon \; \mathbb{N},
\eea

and are chiral since the $s\cdot t = -1$ states vanish \cite{App} (section B). These states are also degenerate, and their number equals the monopole charge $n$. \\




\section{Band structure of Minimal Model} 

In the absence of the magnetic field, the bands for type-I mWSMs ($C \ll v$) are shown as a function of $k_z$ \cite{WSMII} in Fig. 1 (left). They form a continuum with the actual band structure presenting point-like electron pockets at $\mu = 0$ and finite electron pockets at $\mu >0$. For type-II WSMs, the tilt ($C \gg v$) leads to the creation of finite electron and hole pockets at $\mu = 0$, which is the defining characteristic of this phase. These pockets grow unboundedly in the minimal model in the type-I phase as one approaches the Lifshitz transition, and in  the type-II phase, the electron and hole pockets are infinitely large \cite{14,US}. In real type-II mWSMs, higher momentum corrections dominate at large $k$, leading to finite pocket sizes - this is accounted for, in the minimal model, by introducing a momentum cutoff $\Lambda$. \\

The presence of a magnetic field causes the band structure to change dramatically \cite{App} (section C) with one dimensional profiles in the $E-k_z$ planes for each Landau level $N$, as shown in Fig. 2 (left) with the valence and conduction bands being increasingly gapped for $N \geq n$. The violet line depicts the $n$ degenerate zero-modes which are chiral. The $N < n$ LLs have a point-like Fermi surface while the $N \geq n$ LLs have a tube-like Fermi surface \cite{new8} characterized by $N$, and two values of $k_z$, one each for the valence and conduction bands. As the Lifshitz phase transition is approached by tilting the spectrum, the number of LLs which are occupied grows in an unbounded manner, and the minimal model becomes singular at the Lifshitz point, $C = v$. \\

As we move away from the Lifshitz transition into the type-II mWSM regime [Fig. 2 (right)], the Fermi surface still remain point-like for $N < n$ and tube-like for $N \geq n$. However, now there are an infinite number of LLs occupied in the minimal model. Physically, of course, these contributions must be finite since a finite chemical potential cannot support infinite LL occupations. In real WSMs, relevant anharmonic corrections to the Hamiltonian which dominate at larger $\mu$ and only permit the occupation of a finite number of LLs. Since the description provided by the minimal model is valid only for values of $k_z$ and $N$ which are sufficiently close to the Weyl point, we propose that the contribution from the LLs be cutoff at some $N_{max}$ \cite{App} (section G), which is the analog of $\Lambda$ in the $B=0$ case. We define $N_{max} ( \mu )$ as the maximum occupied LL for a given value of the chemical potential $\mu$. $N_{max}$ is constant when $\mu$ lies between two LLs and experiences discontinuous jumps when LLs are crossed by varying $\mu$. \\

\section{Computation of Anomalous Hall conductivity }
We pursue the calculation of zero-frequency DC Hall conductivity tensor in the linear response regime and use the Kubo formula in $d=3$ \cite{App} (section D),

\begin{widetext}
 	\begin{eqnarray}
 	\sigma_{\alpha\beta}^{(s)}(w) &= & - \frac {i}{2\pi \ell_B^2} \sum_{N,N^\prime,t,t^\prime} \int \frac{dk_z}{2\pi} \frac{n_F(E^s_{N,k_z,t})-n_F(E^s_{N^\prime,k_z,t^\prime)}}{E^s_{N,t}-E^s_{N^\prime,t^\prime}} 
 	\frac{\langle N,t,s|J_\alpha|N^\prime,t^\prime,s\rangle  \langle N^\prime,t^\prime,s|J_\beta|N,t,s\rangle}
 	{\omega-E^s_{N,t}+E^s_{N^\prime,t^\prime} - i \eta  }
 	\end{eqnarray},
\end{widetext}

where $J_{\alpha (\beta )}$ represents the current operator, $|N,t,s\rangle$ are the eigenstates, and $\eta_F(E^s_{N,k_z,t})$ is the Fermi-Dirac distribution. For computation of the Hall conductivity we construct the $x$ and $y$ components of the current operators explicitly from the Hamiltonian as,

\begin{widetext}
\begin{eqnarray} \label{jx}
j_x^s=-e \frac{\partial H_{n,B}^s}{\partial (a^\dagger) } \frac{\partial (a^\dagger )}{\partial k_x} -  e \frac{\partial H_{n,B}^s}{\partial (a) }  \frac{\partial (a)}
{\partial k_x}, \:\:\:\:\:\:\:\:\:
j_y^s=-e \frac{\partial H_{n,B}^s}{\partial (a^\dagger) } \frac{\partial (a^\dagger )}{\partial k_y} -  e \frac{\partial H_{n,B}^s}{\partial (a) }  \frac{\partial (a)}
{\partial k_y}
\end{eqnarray}
\end{widetext}

The required matrix elements for the computation are then given by  
\begin{widetext}
 	\bea
 \langle j_{x}^{s}\rangle  &=& -\lambda s n \Big[ C^{\star}_{\uparrow,s,t,k_z,N} C_{\downarrow,s,t^\prime,k_z,N+1}\left( ^NP_{n-1}\right)^{1/2}\delta_{N,N^\prime-1}  + C^{\star}_{\downarrow,s,t,k_z,N^\prime+1} C_{\uparrow,s,t^\prime,k_z,N^\prime}\left( ^{N^\prime}P_{n-1}\right)^{1/2}\delta_{N,N^\prime +1}
 	\Big]\cr
\langle j_{y}^{s}\rangle   &=& i\lambda s n \Big[ C^{\star}_{\uparrow,s,t,k_z,N} C_{\downarrow,s,t^\prime,k_z,N+1}\left( ^NP_{n-1}\right)^{1/2}\delta_{N,N^\prime-1}
  - C^{\star}_{\downarrow,s,t,k_z,N^\prime+1} C_{\uparrow,s,t^\prime,k_z,N^\prime}\left( ^{N^\prime}P_{n-1}\right)^{1/2}\delta_{N,N^\prime +1}
 	\Big] 
\eea
\end{widetext} 

Inserting the expressions for the current matrix elements into the expression for the conductivity tensor, we obtain
\begin{widetext}
\begin{equation}\label{u1}
\sigma_{xy}(\omega=0)   =  -\frac {\lambda^2 n^2}{4\pi \ell_B^2} \sum_{N=n-1}^{N_{max}} {}^NP_{n-1} \sum_{t,t^\prime=\pm}  
 \int_{-\Lambda}^{\Lambda} \frac{dk_z}{2\pi} \frac{n_F(E^s_{N,k_z,t})- n_F(E^s_{N+1,k_z,t^\prime})}{(E^s_{N,k_x,t}-E^s_{N+1,k_z,t^\prime})^2}  \left(1+ \frac{s}{t}\frac{F_0}{F_{N+1}}\right)\left(1- \frac{s}{t^\prime}\frac{F_0}{F_{N+1}}\right),
\end{equation}
\end{widetext} 	
	
where the $F_i$'s are defined below eqn.(\ref{en2}), and $\Lambda$ is the momentum integral cutoff \cite{14,US}. The Mott's rule and the Wiedemann-Franz law \cite{Lutt,14,US} define the  Nernst and thermal Hall conductivities as 

\beq
\alpha_{xy} = eLT \frac{d\sigma_{xy}}{d\mu}, \; \; K_{xy} = LT \sigma_{xy},
\eeq

where $L = \pi^2 k_B^2/3e^2$ is the Lorentz number, $e$ is the electronic charge, and $k_B$ is the Boltzmann constant. Here we wish to note that the vacuum contribution is sourced by the $N=n-1$ term in eqn.(\ref{u1}), since this is precisely where the chiral structure of the ground states come into play, annihilating contributions from zero-modes satisfying $s\cdot t =-1$ (note that these modes don't exist and hence their contribution vanishes). The Hall conductivity tensor can be computed analytically in the $T \rightarrow 0$ and $\omega \rightarrow 0$ (DC) limit, for both type-I and type-II mWSMs, in the tilt-symmetric case $C_+ = -C_- = C >0$.\\

We wish to note that the Hall conductivity being computed here is in the clean limit, i.e., in the absence of impurity scattering. Impurity scattering forms an intrinsic part of any condensed matter system, as no real sample is completely clean. However, it has been argued \cite{new41} that as long as the Fermi surface is sufficiently close to the Weyl nodes, there are no contributions from impurity scattering to the AHE. The inclusion of scattering or to model disorder in the simplest possible way with constant decay widths has been attempted in \cite{gorbar14} with diverging results for the $n=1$ mWSMs, and we expect that higher momentum modes (when $n > 1$) will compound this effect. A more sophisticated approach involves the use of the SBCA (self-consistent Born approximation) technique which has been implemented for the $n=1$ mWSM case \cite{21}. It is shown that in the linear response regime there exists a range of $\mu$ where the effects of scattering can be treated perturbatively and are sub-leading order. While these are interesting avenues of research, we are of the opinion that this is well beyond the scope of the current work. \\

\subsection{Type-I mWSM Hall conductivity.} 
We calculate the type-I AHE conductivity in this section. In this case, after a very lengthy calculation, the computation of the off-diagonal component of the conductivity tensor yields [evaluation of integrals in \cite{App} (section E)],

\begin{widetext}
\begin{equation} \label{w1}
\sigma_{xy}(\omega=0) 
= n \frac{e^2 Q}{2\pi^2} -\frac{e^2}{2\pi^2}\frac{v}{v^2-C^2}
 \left[ n (1-\frac{C}{v})\mu +2\sum_{N=n}^{N_{max}} \sqrt{\mu^2 -\Omega^2(v^2-C^2)^N P_n}  \right]
\end{equation}   
\end{widetext}



with $N_{max}$ set by the chemical potential $\mu$ (assumed to be positive, without loss of generality), and tilt as ${^N P_n} = \frac{\mu^2}{(v^2-C^2)\Omega^2}$, with $N_{max} = \lfloor N \rfloor$, where we have assumed that the chemical potential lies between two Landau levels \cite{26}. For $n=2$, we find that $N^{n=2}_{max} (\mu) = \lfloor \frac{1}{2}+ \sqrt{\frac{1}{4} + \frac{\mu^2}{(v^2-C^2)\Omega^2}} \rfloor$. Similarly, for $n=3$ the relationship yeilds $ N^{n=3}_{max} (\mu) = \lfloor 1+ (\frac{L}{18})^{1/3}+ (\frac{2}{3L})^{1/3} \rfloor $, where $L \equiv \sqrt{\frac{81\mu^4}{4(v^2-C^2)^2\Omega^4} - 12} + \frac{9\mu^2}{(v^2-C^2)\Omega^2}$. As expected, the result in eqn.(\ref{w1}) is a very natural generalization of the $n=1$ case \cite{21,gorbar14}, with $C \rightarrow 0$. For a generic monopole charge and $\mu \neq 0$ there are no closed form solutions for $n \geq 5$, but one can show that there is only one physical solution $N_{max}$, for $\mu > 0, \forall n$ \cite{App} (section F). \\

Note that algebraically we find that the only ground state that contributes to the current correlation function is $N=n-1$. However, it contributes with strength $n$, which is the monopole charge and the number of degenerate ground states, and hence this contribution can be interpreted as all $n$ ground states contributing equally.  \\

\begin{widetext}

\begin{figure}[h]
\centering
\fbox{\includegraphics[scale = .6]{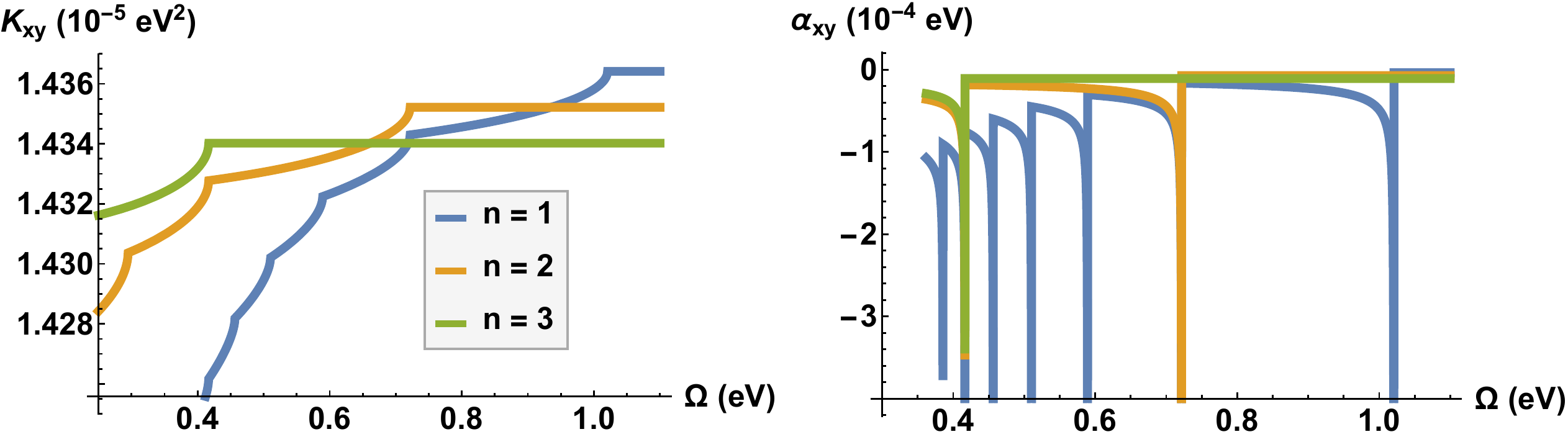}}
\caption{Left panel: The thermal Hall conductivity of a type-I mWSM, plotted as a function of magnetic field dependent LL spacing $\Omega$ with units of momentum. Right panel: The Nernst conductivity of a type-I mWSM, plotted as a function of magnetic field B dependent LL spacing $\Omega$. The plots have been shown for $n=1,2,3$ and are at temperature $T=10^{-3} K$. The values of the other parameters for both plots are: $Q = 1000$eV, $\mu = 0.001$ eV, $v = 0.001$, $C = 0.0002$.}
\end{figure}	
	
\begin{figure}[h]
\centering
\fbox{\includegraphics[scale = .5]{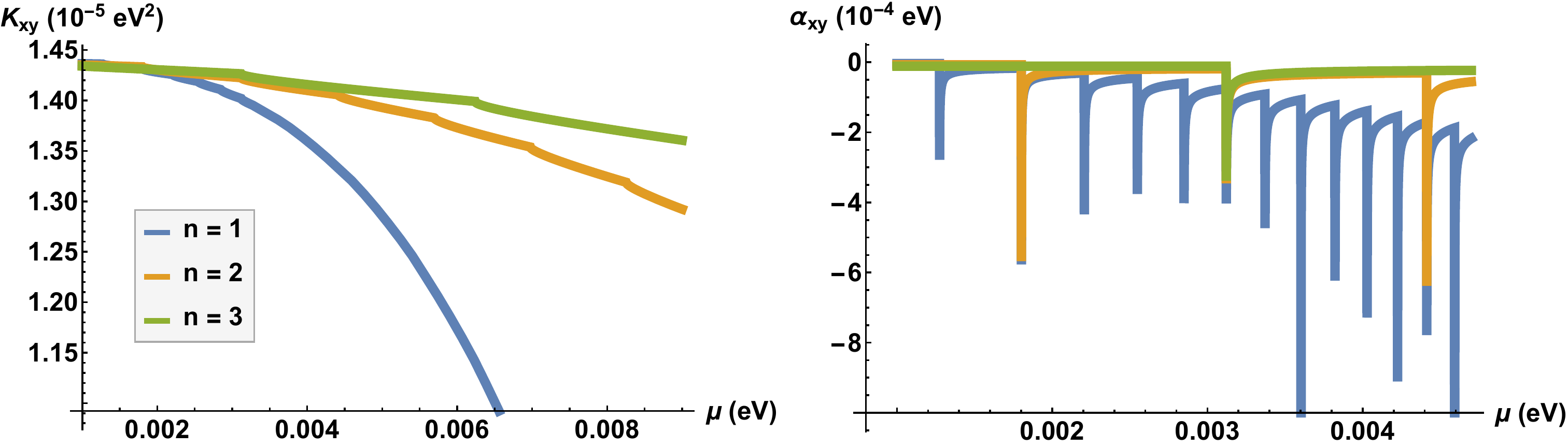}}
\caption{Left panel: The thermal Hall conductivity of a type-I mWSM, plotted as a function of chemical potential $\mu$ with units of momentum. Right panel: The Nernst conductivity of a type-I mWSM, plotted as a function of chemical potential $\mu$. The plots have been shown for $n=1,2,3$ and are at temperature $T=10^{-3} K$. The values of the other parameters for both plots are: $Q = 1000$eV, $\Omega = 1.3$ eV, $v = 0.001$, $C = 0.0002$.}
\end{figure}
\end{widetext}

The expression for $\sigma_{xy}$ becomes singular as $C \rightarrow v$, a consequence of the appearance of infinite LL contributions in the minimal model when approaching the Lifshitz transition from the mWSM-I side. As in the zero-field elementary WSM case, the expression for the $\sigma_{xy}$ is independent of the momentum cutoff $\Lambda$. The thermal Hall and Nernst conductivities are plotted as a function of LL spacing $\Omega$ [Fig. 3] and chemical potential [Fig. 4], using Natural units. The results obtained in Fig. 3(a) reproduces the features in Fig. 4 of \cite{gorbar14} for $n=1$. One can understand the Hall conductivity in terms of the volume of electron and hole pockets for a given Fermi energy in the zero temperature limit. For the type-I phase one observes purely electron pockets and so the Hall conductivity decreases as a function of $\mu$, due to a growth in the size of the pockets. This is precisely the behavior shown in Fig. 4(a), and the reason that $\sigma_{xy}$ starts off at a positive value is the topological vacuum contribution. The Nernst conductivity, being related to the Hall conductivity via a derivative, experiences discountinuous jumps as shown in Fig. 4(b) as $\mu$ is increased leading to LL crossings. Since the Nernst conductivity is coupled to the derivative of $\sigma_{xy}$, it peaks sharply at the crossing points of LLs. Similar effects are observed when $\Omega$ is varied which affects the LL spacing and hence the number of occupied states for a given $\mu$. \\

\subsection{Type-II mWSM Hall conductivity} The results for type-II mWSMs are as follows \cite{App} (section G):

\begin{widetext}
	
\bea \label{ab30}
\sigma_{xy}(\omega=0) =  - \frac{e^2}{2 \pi^2} \left[ { \frac{2 \mu C}{C^2-v^2} (N_{max}+1) + \frac{n \mu}{C-v}- \frac{2}{v} \frac{C^2+v^2}{C^2-v^2} n \mu} \right]
\eea 

\end{widetext} 

\begin{widetext}

\begin{figure}[h]
	\centering

		\fbox{\includegraphics[scale = .48]{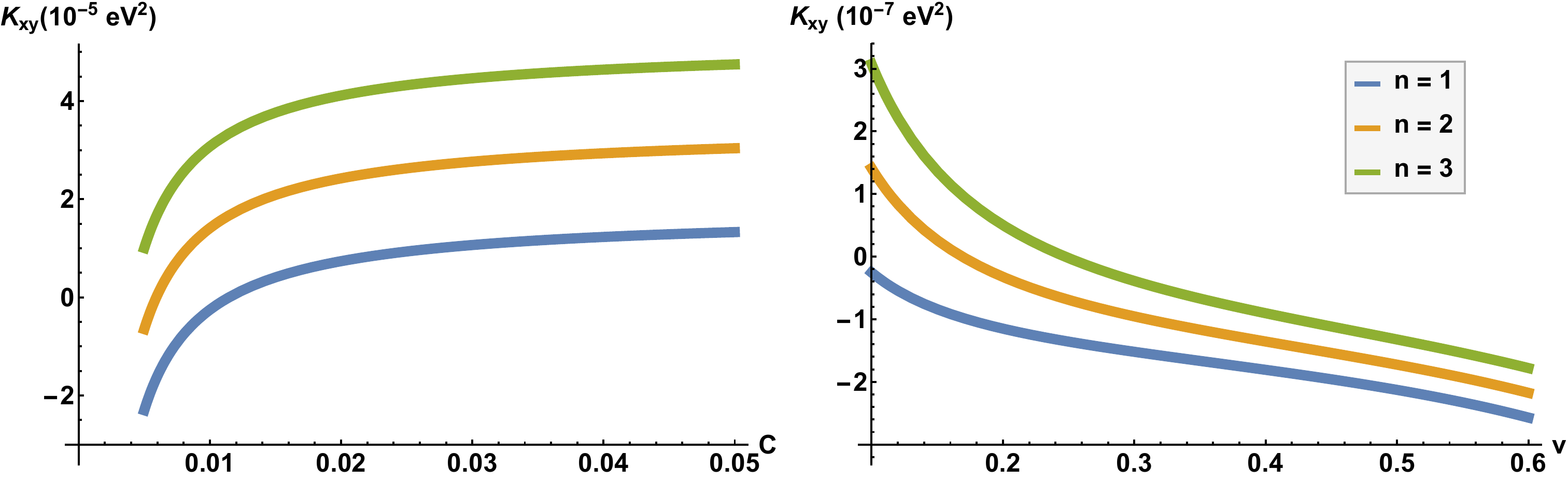}}
		 \caption{Left panel: The thermal Hall conductivity for a type-II mWSM is plotted as a function of tilt $C$, at fixed $v$. The values of the other parameters are:  $\mu = 0.6$ eV, $v = 0.001$, $N_{\text{max}} = 10$. Right panel: The thermal Hall conductivity for a type-II mWSM is plotted as a function of Fermi velocity $v$, at fixed tilt. The values of the other parameters are:  $\mu = 0.6$ eV, $C = 1$, $N_{\text{max}} = 10$. Both plots include the different curves for $n=1,2,3$, and have $T=10^{-3}K$.}
	\end{figure}
	
	
\end{widetext}

Note that the anomalous node term has vanished, with $N_{max}$ being the type-II phase cutoff as discussed previously. By viewing the magnetic band structure as discrete sections of the continuum band structure, one can estimate $N_{max}$. The continuum model cutoffs $\Lambda_x(\mu)$, $\Lambda_y(\mu)$, and $\Lambda_z(\mu)$ can be calculated from the first principles band structure \cite{FP1} for the type-II class of mWSMs, corresponding to points where the higher momentum corrections force the bands to cross the Fermi surface away from the Weyl points - these values then depend on $\mu$. Comparing eqn.(\ref{CB1}) and eqn.(\ref{en2}), we see that $\gamma^2 (k_x^2 + k_y^2)^n \sim ~^N P_n \Omega^2$. And so $N_{max}$, in the type-II phase, can be obtained as $ ~^{N_{max}} P_n \sim \lfloor \gamma^2 (\Lambda_x^2 + \Lambda_y^2)^n/\Omega^2 \rfloor$, where $\lfloor N \rfloor$ restricts $N$ to the nearest lower integer. \\

The Hall conductivity grows linearly with increasing $\mu$, leading to a constant Nernst coefficient as long as new LLs are not crossed. The variation of thermal Hall with tilt and Fermi velocity are depicted in Fig. 5. By examining eqn.(\ref{ab30}) we see that there are competing positive and negative $v$ and $C$ dependent terms and hence $K_{xy}$ can be positive or negative, depending on the values of these parameters. In the type-II phase, one of the novel characteristics is the presence of hole pockets. Both types of carriers coexist and compete, and in fact there is a cross-over when the tilt is sufficiently large, such that the hole pockets dominate the Hall physics leading to a positive Hall conductivity as shown in Fig. 5(a). The value of tilt at which the Hall conductivity changes sign in our model can be calcuated \cite{App} (section H) from the expression above as  

\beq
C_{crit}^{+} = \frac{v}{4n} \left[ (n+2N_{max}+2) \pm \sqrt{(n+2N_{max}+2)^2 -8n^2}\right] \nonumber
\eeq

Since $\sigma_{xy}$ is linear in $\mu$ in the type-II regime, we see that $\alpha_{xy} = eLT \frac{d\sigma_{xy}}{d\mu} = eLT \frac{\sigma_{xy}}{\mu} = \frac{e}{\mu} K_{xy}$. This shows that $\alpha_{xy}$ is proportional to $K_{xy}$ when $\mu$ is held fixed, indicating that the two quantities share similar qualitative features. \\

The anisotropic dispersion of a general mWSM close to the Weyl node in the presence of a perpendicular magnetic field can be confirmed using angle-resolved photo-emission spectroscopy \cite{new9,arp3,arp4,arp5,arp6,arp7}. Also, a simple closed-circuit setup \cite{14,US} can be used to verify the claims made in this manuscript, i.e., eqns. (\ref{w1}) \& (\ref{ab30}). The Nernst and thermal Hall coefficients calculated here represent the system's response to a current in the linear regime, and can be measured directly.  \\

\section{Vanishing vacuum contribution of AHE in type-II \lowercase{m}WSM\lowercase{s}}

\subsection{T $\rightarrow$ 0 limit vacuum AHE }

We now analyze the vanishing of the AHE vacuum contribution in the type-II mWSM phase. We seek to understand this phenomenon from the perspective of band structure and occupations in the $T \rightarrow 0$ limit. Conductivity can be understood in terms of scattering between states and in this context the vacuum contribution to the AHE comes from the scattering of the vacuum state to higher LLs. For the minimal model, the only non-vanishing matrix element is between the $n$ degenerate vacuum states and the next non-degenerate LL. \\

\begin{figure} [h]
	\centering
	\begin{subfigure}[t]{.45\textwidth}
		\centering
		\fbox{\includegraphics[scale = .5]{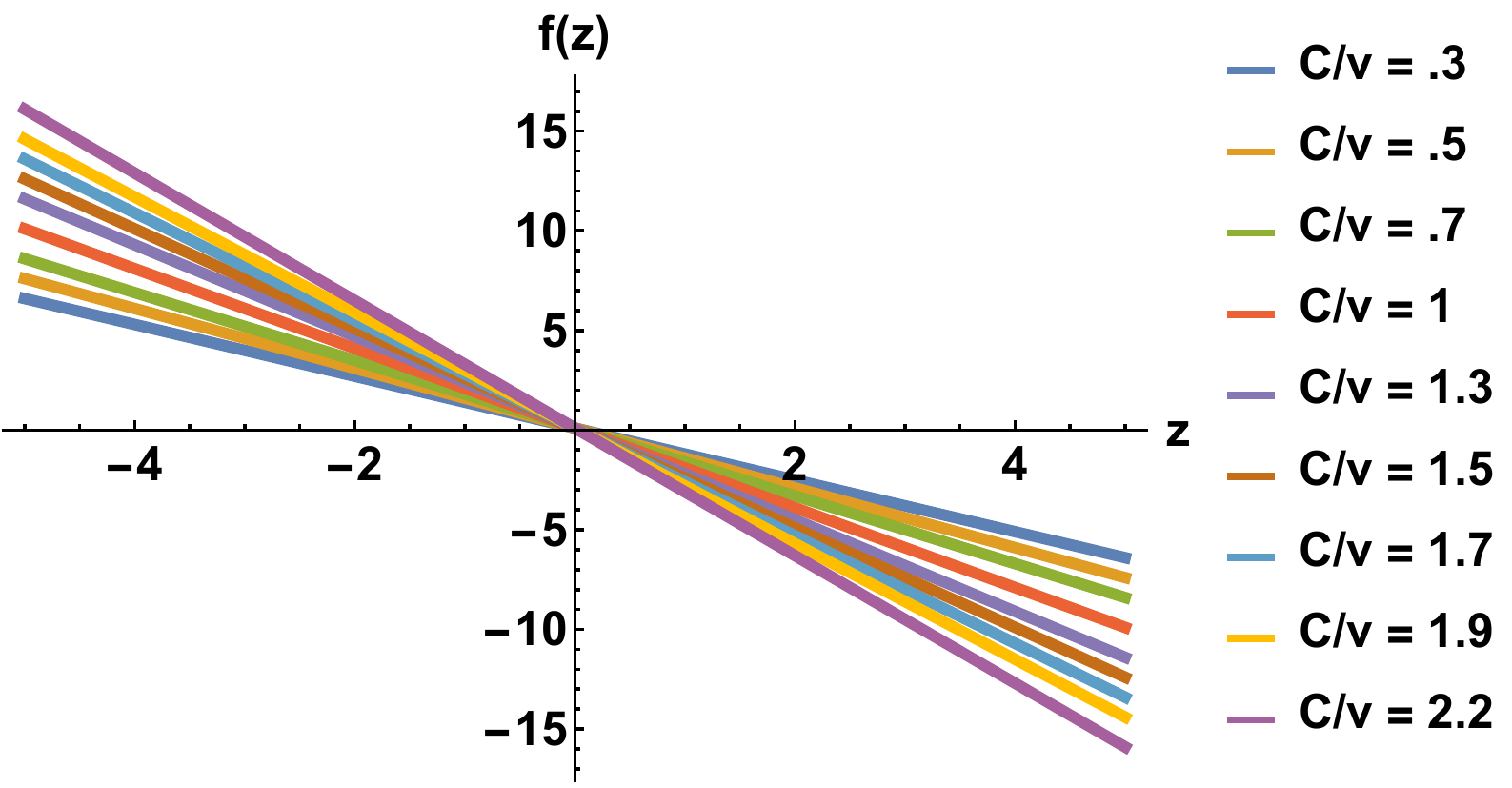}}
		\caption{} 
	\end{subfigure}
	\begin{subfigure}[t]{.45\textwidth}
		\centering
		\fbox{\includegraphics[scale = .5]{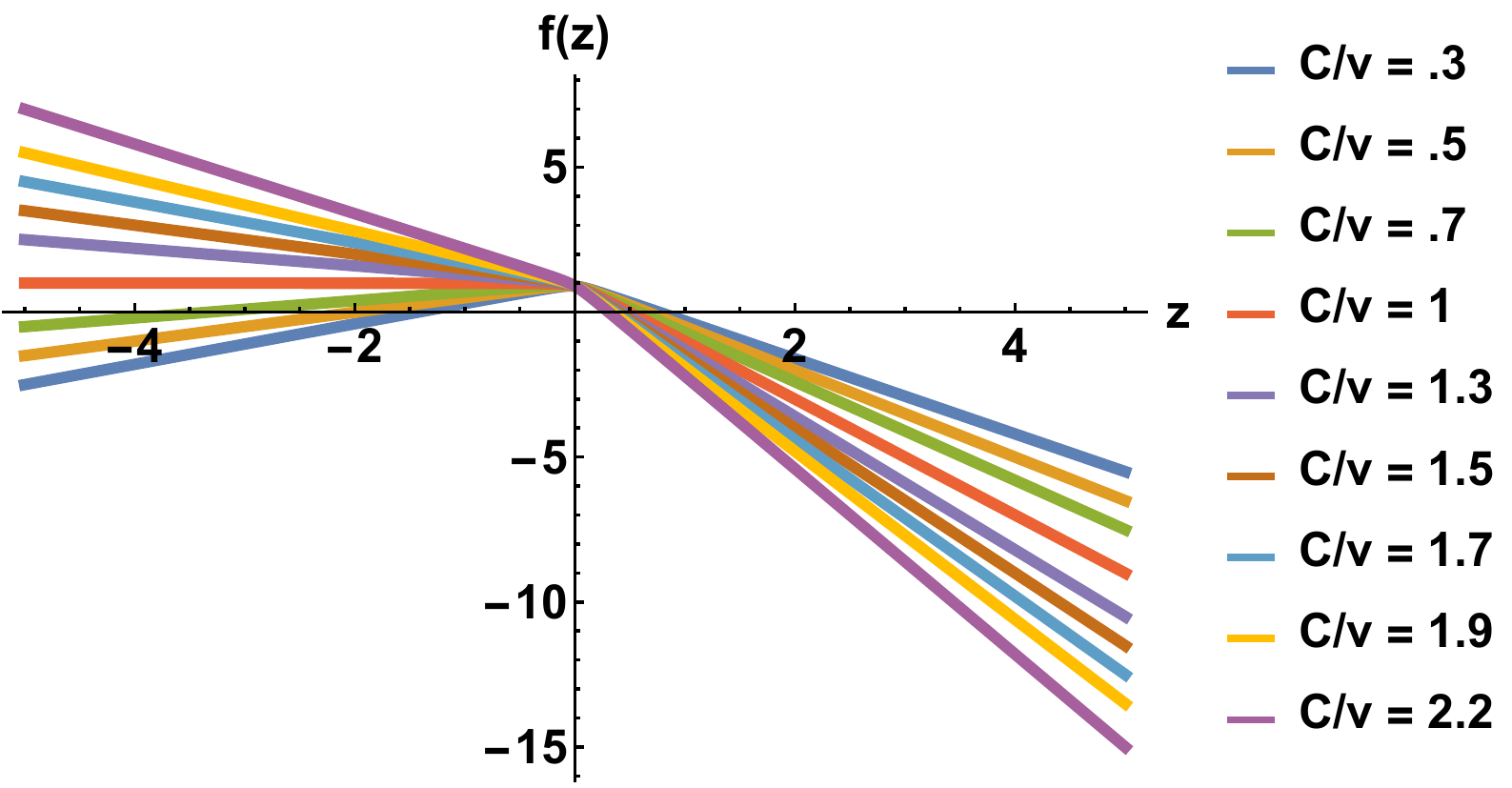}}
		\caption{} 
	\end{subfigure}

	\begin{subfigure}[t]{.45\textwidth}
		\centering
		\fbox{\includegraphics[scale = .5]{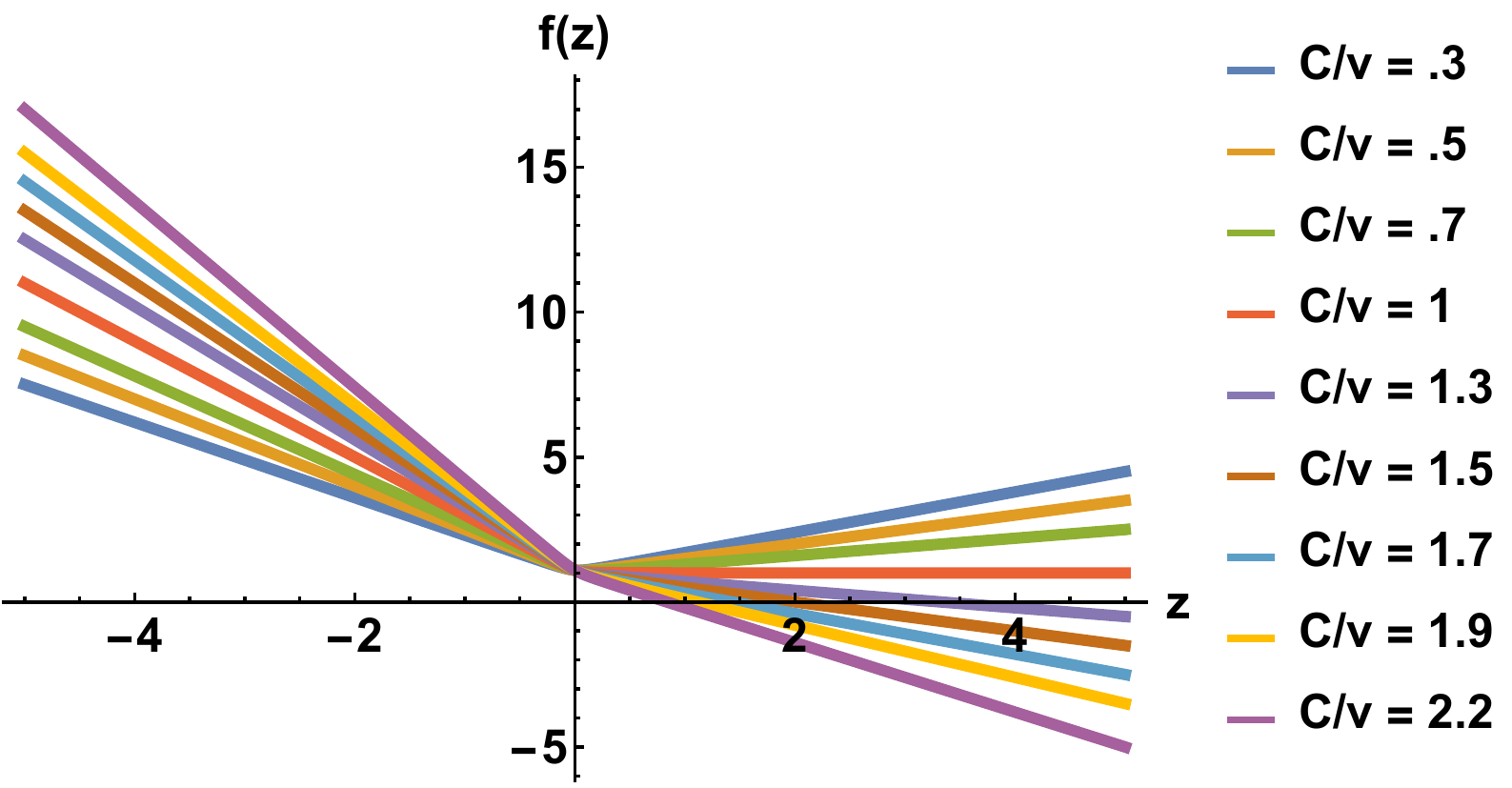}}
		\caption{} 
	\end{subfigure}

\caption{The values shown in the legend are those of $\alpha = C/v$ indicating the relevant phase. (a) The inequality function for the ground states (b) The inequality function for the $N \geq n$ states $t>0$ (c) The inequality function for the $N \geq n$ states $t<0$ 	}
\end{figure}

In eqn.(\ref{u1}), the AHE vacuum contribution is proportional to the difference in occupations of the vacuum states and the $N = n$ LL. The occupation of states is governed by the Fermi-Dirac (FD) distribution which is defined as $n_F(E) = [\exp(\frac{E - \mu}{T}) + 1]^{-1}$, where $E$ is the energy of the state, $\mu$ is the chemical potential, and $T$ is the temperature. In the zero temperature limit the FD distribution reduces to a Heaviside step functions of the form $\Theta [\mu - E_{N,s,t}(k_z)]$, where the energies are defined in eqn.(\ref{en2}). A state potentially contributes to the vacuum AHE when it is occupied, i.e., $\Theta [\mu - E_{N,s,t}(k_z)] = 1$. This condition is satisfied when the argument of the Heaviside function is positive which is stated as: 

\begin{equation} \label{ee}
\mu^{'} - \alpha z \pm \sqrt{z^2 + \Delta^2} \geq 0,
\end{equation}

for $s = +$ and $t = \pm$. We consider the effects of tilting on the $s=+$ node and the results turn out to be identical for the other node. $\Delta = {}^NP_n \Omega^2$ models the Landau level (LL) contribution and separates the inequality functions into two equivalence classes modulo the value of $\Delta$: the class of excited states has $\Delta \neq 0$ and the ground states have $\Delta = 0$. $z = k_z - Q$ is the shifted momentum in the $z$ direction. The parameter $\alpha$ is the ratio $C/v$ which defines the two phases - if $\alpha \ll (\gg) 1$ we are in the heart of the type-I (II) phase. $\mu^{'} = \mu /v$ is the rescaled Fermi energy, defined for convenience. For the ground state we have that $\Delta = 0$ and this operation picks out the negative branch from eqn.(\ref{ee}) since for $s=+$, only the $t=+$ chiral ground state exists.\\  

We define the expression on the lhs of eqn.(\ref{ee}) as $f(z) = \mu^{'} - \alpha z \pm \sqrt{z^2 + \Delta^2}$. The plots for $f(z)$ are presented for the excited states of both mWSM phases in Fig. 6. In Fig. 6(a) we plot $f(z)$ as a function of $z$ for the chiral ground states, in Fig. 6(b) we plot the $N \geq n$ states of mWSMs with $t>0$, and in Fig. 6(c) we plot the $N \geq n$ states with $t<0$. These plots indicate the values of $z$ for which the inequalities in eqn.(\ref{ee}) are satisfied, and the arguments in what follows are based on them. \\

In order for a state to contribute to the AHE, the inequality in eqn.(\ref{ee}) has to hold, and the node separation $Q$ (i.e. the topological vacuum term) only contributes when the inequality satisfying range includes the cutoff $\Lambda$. In the type-I ($\alpha < 1$) phase the inequality function for $t=-$ is always positive and goes to positive infinity at large $|z|$ leading to an overall contribution of $2Q$ to the AHE, up to prefactors. The ground state inequality for both phases contribute zero, despite the inequality being satisfied at large $|z|$ for $z<0$, because the ground state couples with opposite sign to the $N = n, t = \pm$ states. This leads to the effective node dependence $2Q$ as obtained in eqn.(E27) of \cite{App}. The $t=+$ inequality function for $\Delta > 0$ is bounded for positive values and does not contribute to the vacuum AHE. For the type-II phase ($\alpha > 1$), the non-zero mode inequalities are positive at large $|z|$ only for $z<0$, and so each of them only contribute one factor of $Q$. Crucially, they contribute with opposite sign, and this leads to the annihilation of the vacuum contribution \cite{App} (section G). \\ 

The key point lies in the domination of the tilt in the type-II phase which leads to the cancellation of the occupations at the cutoff, leading to the vacuum AHE being both cutoff and node-independent, i.e., zero. One can understand this in terms of the hole pockets which arise in the type-II phase \cite{WSMII}, which in the absence of the strong magnetic field lead to a decreased and non-universal vacuum contribution due to hole and electron contributions competing \cite{zyuzin16}. In the presence of the magnetic field this destructive interference is absolute! One can trace this back to the idea that in the absence of the magnetic field, there are extended hole and electron pockets in the spectrum of the type-II mWSM whose relative sizes depend on the tilt angle. However, in the presence of the magnetic field, the pockets are tube-like and their sizes are characterized by $N$, and so the $N=n$ holes and electron contributions cancel exactly.\\

\subsection{Anomalous vacuum contribution at all temperatures}

We examine the anomalous part of vacuum contribution at arbitrary temperature for the type-II mWSM phase. We begin with eqn.(\ref{u1}) and find a compact form [details in \cite{App} section I)] as shown below.
\begin{widetext}
\begin{eqnarray}\label{u8}
\sigma_{xy}^{(s)}(\omega=0) & =& -\frac{\lambda^2}{8\pi \ell_B^2} \frac{1}{\Omega^2v^2} \sum_{N=n-1}^{N_{max}} (2N+2-n) \sum_{t,t^\prime=\pm} 
\int_{-\infty}^{\infty} dk_z \{ n_F(E^s_{N,k_z,t})- n_F(E^s_{N+1,k_z,t^\prime})\} \\ \nonumber 
& - & \frac{\lambda^2}{8\pi \ell_B^2} \frac{n}{\Omega^2v^2} s \sum_{N=n-1}^{N_{max}} \sum_{t,t^\prime=\pm} \int_{-\infty}^{\infty} dk_z 
\{ \frac{n_F(E^s_{N,k_z,t)}}{tF_N} - \frac{n_F(E^s_{N+1,k_z,t^\prime})}{t^\prime F_{N+1}}\} F_0
\end{eqnarray}
\end{widetext}

To obtain the vacuum contribution, we start with the expression (\ref{u8}) and examine the $N = n-1$ term. The non-zero terms are, up to overall prefactor $-\frac{\lambda^2}{8\pi \ell_B^2} \frac{4n}{\Omega^2v^2}$, shown below. 
\begin{widetext}
\begin{eqnarray}\label{u11}
 \sigma_{xy}^{(s=+)}(\omega=0)  &=  \int_{-\infty}^{\infty} dk_z 
\left[ 2n_F(E^+_{n-1,k_z,+})-  n_F(E^+_{n,k_z,+}) - n_F(E^+_{n,k_z,-}) - n_F(E^+_{n,k_z,+})\frac{F_0}{F_n} +  n_F( E^+_{n,k_z,-})\frac{F_0}{F_n} \right]
\nonumber \\
 \sigma_{xy}^{(s=-)}(\omega=0)  &=\int_{-\infty}^{\infty} dk_z 
\left[ 2n_F(E^-_{n-1,k_z,-})-  n_F(E^-_{n,k_z,+}) - n_F(E^-_{n,k_z,-}) + n_F(E^-_{n,k_z,+})\frac{F_0}{F_n} -  n_F( E^-_{n,k_z,-})\frac{F_0}{F_n} \right]
\end{eqnarray}
\end{widetext}

The zero-mode contribution in eqn. (\ref{u11}) contains the so called {\it anomalous} vacuum contribution, the part that exists at $\mu=0$. We are not interested in the other $\mu$ dependent terms for now - we want to extract the behavior of the anomalous term for $T\neq 0$. For $T=0$ we know that this term is proportional to $Q$ for type-I mWSMs and zero for type-II mWSMs. Similar to \cite{gorbar14}, we obtain the anomalous contribution which reads (with $\Lambda \rightarrow \infty$)
\begin{widetext}
\begin{eqnarray}\label{u13}
\sigma_{xy}^{(s)} \mid^{\textrm{anom}}  = \frac{\lambda^2}{8\pi \ell_B^2} \frac{4n}{\Omega^2v^2} \int_{-\Lambda}^{\Lambda} dk_z
 \left[- n_F(E^+_{n,k_z, +}) +  n_F(E^+_{n,k_z,-})
 +  n_F(E^-_{n,k_z, +}) - n_F(E^-_{n,k_z,-}) \right] \frac{F_0}{F_n}
\end{eqnarray}
\end{widetext}
 
One can check that this gives the anomalous contribution for $T\rightarrow 0$, and that the other terms do not contribute \cite{App} (section I). In the presence of tilt, the integrals in eqn.(\ref{u13}) are not analytically tractable. To proceed, one needs to add cutoff dependent terms ($\chi$) proportional to $C$ to the integral for each of the Fermi functions [These terms needs to then be subtracted off appropriately after integration to obtain finite results \cite{App} (sections I \& J)]. Applying the technique described in detail in \cite{App}, the anomalous vacuum contribution in the expression of the Hall conductivity tensor for mWSMs reads,

\begin{widetext}
\begin{eqnarray}\label{u17}
\sigma^{anom}_{xy}(\omega=0) + \chi = - \frac{\lambda^2}{2\pi \ell_B^2}\frac{n}{\Omega^2v^3 \beta}
\left[
\begin{aligned}
  & \ln  \Bigg{(}
\frac{1+e^{-\beta(C(k_z-Q) +v\sqrt{(k_z-Q)^2+\Omega^2 {}^NP_n}-\mu} } {1+e^{-\beta(C(k_z-Q) -v\sqrt{(k_z-Q)^2+\Omega^2 {}^NP_n}-\mu} } \Bigg{)} 
\end{aligned} \right.{}
\nonumber \\
\left.{}
\begin{aligned}
  +   & \ln  \Bigg{(}
\frac{ 1+e^{-\beta(C(k_z +Q) +v\sqrt{(k_z +Q)^2+\Omega^2 {}^NP_n}-\mu} } {1+e^{-\beta(C(k_z +Q) -v\sqrt{(k_z +Q)^2+\Omega^2 {}^NP_n}-\mu} } \Bigg{)}
\end{aligned}
\right]_{-\Lambda}^{\Lambda}.
\end{eqnarray}
\end{widetext} 

Now we can find out the expression for the anomalous vacuum contribution for the case of type-I and type-II phases separately. For type-II mWSM, one finds that, 
\begin{equation}\label{u18}
\lim_{\Lambda \rightarrow \infty} \sigma^{anom}_{xy}(\omega=0) \Big\lvert_{\textrm{type-II}} =0.
\end{equation}
This makes the vacuum contribution zero for all temperatures  $T$ and linear response regime values of $\mu$! For the type-I phase, the expression for the anomalous Hall conductivity tensor has to be evaluated very carefully. Using eqn (\ref{u17}), a detailed and tedious calculation with proper evaluation of the integrals leads us to find that for the type-I phase $\lim_{\Lambda \rightarrow \infty} \sigma^{anom}_{xy}(\omega=0) + \chi \lvert_{\textrm{type-I}} = n \frac{e^2}{2 \pi^2} \left(Q +\frac{C}{v} \Lambda\right)$. This is not the final expression for the AHC for the type-I phase in mWSMs. It may be noted that this expression contains a term proportional to $\frac{C}{v}\Lambda$. The integration picks out the correct contribution but presents an additional cutoff dependent term, due to the added term $\chi$, and it becomes clear why the integral is analytically tractable in the $C=0$ case. Subtracting off the cutoff dependent term after integration \cite{App} (sections I \& J) yields  

\begin{equation}\label{u20}
\lim_{\Lambda \rightarrow \infty} \sigma^{anom}_{xy}(\omega=0) \Big\lvert_{\textrm{type-I}} = n \frac{e^2}{2 \pi^2} Q .
\end{equation}

The multiplicative factor $n$ shows the dependence of the monopole charge on the AHC in type-I phase. The results show that the qualitative and quantitative difference in the behaviour of the vacuum AHC for the type-I and type-II phase of of mWSM. It is zero for the type-II phase for all temperatures and non-zero for the type-I phase.\\

\section{Conclusion} 

In this paper, we have examined the effects of a perpendicular magnetic field on generic type-I and type-II multi-Weyl Semimetals in their respective minimal models, in the high magnetic field limit or Landau regime. We have analyzed the structure of the Hilbert space, and computed the Hall conductivity tensor in the linear response regime using the Kubo formula in the zero-frequency limit.\\

We find that the Hilbert space, in the presence of the quantizing field, hosts $n$ zero modes which are both degenerate and chiral, while the higher Landau levels are achiral. The chiral structure of the zero-modes that contributes to the anomalous Hall conductivity leads to the preservation of the topological node separation ($Q$) term in the Hall conductivity for the type-I phase. The type-I Fermi-surface correction to anomalous Hall conductivity generalize naturally from the $n=1$ case \cite{21,gorbar14}. \\

For type-II WSMs the Hall component of the conductivity tensor gets contributions from all of the countably infinite LLs, which is a feature of the minimal model in the $B\neq 0$ case. Since a real material will only host a finite number of occupied LLs for a given chemical potential, we introduce a LL cutoff $N_{max}$, similar in spirit to the standard momentum cutoff introduced in the $B=0$ case. We make a crude estimation of this cutoff in terms of model parameters. The Nernst conductivity in this model remains a constant for varying $\mu$ unless a LL is crossed - here $N_{max}$ presents a discrete jump and the Nernst conductivity does the same. Interestingly, we find that the anomalous vacuum contribution in type-II mWSMs is annihilated in the presence of the magnetic field at zero temperature. We understand this in terms of the cancellation of electron and hole pockets which are tube-like in the presence of the perpendicular $B$ field. This leads us to examine the anomalous vacuum contribution at all temperatures,  and we find that it does indeed vanish at all temperatures. These results will be tested in a more realistic two-band model in a future work. \\

The qualitative and quantitative observations made in this manuscript are designed to serve in the characterization of generic mWSMs of both types, putting their properties in an equivalence class modulo the two types of tilt. \\

{\textit {\textbf{Acknowledgement}:}} The authors would like to thank Sergey Y. Savrasov and Giacomo Resta for enlightening conversations and constructive comments regarding the work presented in this manuscript. \\

\end{document}